%% file: bare_jrnl.tex
\documentclass[conference]{IEEEtran}
\IEEEoverridecommandlockouts
\usepackage{url}

\usepackage{adjustbox}
\usepackage{amsmath,graphicx}
\usepackage{xcolor}
\usepackage[normalem]{ulem}
\usepackage{url}
\usepackage{multirow}
\usepackage{placeins}
\usepackage{enumitem}
\usepackage[nomain, acronym, symbols]{glossaries}
\usepackage{siunitx}
\usepackage{bm}

\newcommand{\matr}[1]{\mathbf{#1}}     
\renewcommand{\vec}[1]{\mathbf{#1}}

\input{other/acronyms}

\def\BibTeX{{\rm B\kern-.05em{\sc i\kern-.025em b}\kern-.08em
    T\kern-.1667em\lower.7ex\hbox{E}\kern-.125emX}}
\begin{document}

\title{Geometry-Based Compression of Plenoptic Point Clouds\\
\thanks{Work partially supported by CNPq under grant 301647/2018-6; 
This project has received funding from the European Union’s Horizon 2020 research and innovation
programme under the Marie Sk\l odowska-Curie grant agreement No 956770.}
}


\author{\IEEEauthorblockN{Davi R. Freitas}
\IEEEauthorblockA{\textit{Inria - Rennes Bretagne-Atlantique}\\
Rennes, France \\
rabbouni.davi@ieee.org}
\and
\IEEEauthorblockN{Gustavo L. Sandri}
\IEEEauthorblockA{\textit{Instituto Federal de Brasília}\\
Brasília, Brazil \\
gustavo.sandri@ieee.org}
\and
\IEEEauthorblockN{Ricardo L. de Queiroz}
\IEEEauthorblockA{\textit{Universidade de Brasília} \\
Brasília, Brazil \\
queiroz@ieee.org}
}

\maketitle

\begin{abstract}
Plenoptic point clouds (PPC) are novel data structures that represent the light from different viewing directions in order to provide a higher degree of realism to regular point clouds. This is achieved by associating each point to multiple colors instead of a single one. 
Here, we present a method to efficiently compress the attributes of a PPC, consisting of a Karhunen-Lo\`eve transform over the color attributes followed by multiple attribute coders with intra prediction capability. This compression scheme can be incorporated within the MPEG’s geometry-based PCC (G-PCC) standard, using any of G-PCC’s existing solutions for attribute coding. Compression performance assessment using PPCs of different spatial resolutions reveals competitive results in comparison to existing methods, such as RAHT-based or video-based PCC solutions. We believe our coder to be the new state of the art.
\end{abstract}



\section{Introduction}


A \gls{pc} is a data structure that represents objects and space in a \gls{3d} coordinate system. Frequently used in applications of real-time rendering and capture of \gls{3d} objects \cite{use_cases}, it is common to voxelize points by constraining them to a grid of $ N\times N\times N $ voxels \cite{voxelized_pc}. 
\gls{pc}s have been favored over meshes due to the greater simplicity in its capture process. However, there is a huge amount of data required to represent these data structures.  Therefore, the development of compression techniques in order to transmit and store this data is paramount. As a result, the \gls{mpeg} is among groups with ongoing standardization processes towards point cloud data compression \cite{emerging_mpeg}.

The \gls{gpcc} standard has been proposed by \gls{mpeg} to compress high-detail static point clouds as efficiently as possible \cite{emerging_mpeg, overview_gpcc_vpcc, gpcc_codec_description}. \gls{gpcc} offers two distinct algorithms for attribute coding: an encoder based on the \gls{raht} \cite{raht} and one based on \gls{lod}.

The plenoptic function aims to represent the intensity of light at any given point by a 7-dimensional function \begin{equation}\label{fun:pleno}
    P(x,y,z,\theta,\phi,\lambda,t),
\end{equation} 
where $(x,y,z)$ are space coordinates, $(\theta,\phi)$ indicate the viewing direction, $\lambda$ is the light wavelength and $t$ is the time \cite{plenoptic_func}. A \gls{ppc} can represent such a function by discretizing the parameters. The coordinates $(x,y,z)$ are discretized into voxel positions, the wavelength $\lambda$ is discretized into \gls{rgb} color components and time $t$ is discretized with the capture of successive frames. The azimuth and elevation angles $(\theta,\phi)$ are naturally sampled by a finite number of camera rigs in the capture process, and the intermediate values for these parameters can be interpolated by the renderer. Hence, a dynamic point cloud with $N_c$ color attributes per voxel --- where $N_c$ represents the number of camera rigs --- can represent the plenoptic function and it is referred here as a \gls{ppc}.

Point cloud compression is an active research field \cite{emerging_mpeg} and the representation of the plenoptic function has been the subject of MPEG and JPEG's ongoing standardization activities \cite{jpeg_pleno, mpeg_pleno}. Sandri et al. proposed the compression of \gls{ppc}s by several approaches using the \gls{raht} coder \cite{sandri_icip, sandri_ppc}. The results have shown that the method using the \gls{klt} over the voxel colors, followed by the use of \gls{raht} over the transformed attributes, achieved the best performance at the time. Later on, Krivoku\'{c}a et al. proposed subdividing the \gls{ppc} into clusters of specular and diffuse components prior to \gls{klt} and \gls{raht} \cite{raht_klt_clusters}. Zhang et al. model the \gls{ppc} from a different perspective, continuously representing the color function over $(\theta,\phi)$ and compressing the spherical functions \cite{zhang}. Naik et al. proposed a solution to handle \gls{ppc} data with MPEG's \gls{vpcc} standard \cite{nokia_1}. Afterwards, they proposed optimizations to this scheme by discarding some of the views based on the voxels' specularity \cite{nokia_2}. Li et al. proposed a video-based solution using \gls{vpcc} by compressing the multiple attributes via the \gls{mv-hevc} \cite{lili}. Finally, Krivoku\'{c}a et al. \cite{maja_6d} introduces a representation of \gls{ppc}s as sets of 6-D spatio-angular locations, providing an occlusion-aware compression method for the plenoptic attributes.

These solutions, however, either are not competitive with the existing state-of-the-art, or are not compliant with the \gls{gpcc} standard. On the other hand, our proposal is capable of enhancing TMC13's framework with plenoptic features while being competitive with existing leading methods.

\section{Proposed Framework} \label{sec:framework}


The demand of representing 3D scenes within MPEG'S standards in a realistic and compact manner led us to develop an approach to compress a \gls{pc} with plenoptic features within their test model.

Thus, consider a \gls{ppc} where the $n$-th occupied voxel has a set of $N_c$ \gls{rgb} color values associated to the different camera viewpoints used in the capture process. Assume the \gls{rgb} values are converted to YUV space, yielding plenoptic color vectors $[Y_{n}^{1}, U_{n}^{1}, V_{n}^{1}, \dotsm , Y_{n}^{N_c}, U_{n}^{N_c}, V_{n}^{N_c}]$. This is due to the fact that the camera color variation is most prominent in the luminance component, which best reflects the changes in specularity. Thus, the smaller variance in the chrominance channels is a property that can be explored to enhance the compression. Hence, we apply a linear transform over each of the color components Y, U and V in order to take advantage of the correlation between the $N_{c}$ plenoptic views. Consider $C$ as one of the color channels and let

\begin{equation}\label{fun:color-comp}
    \vec{c}(n) = [C_{n}^{1} , C_{n}^{2} , C_{n}^{3} , \dotsm , C_{n}^{N_c}]^T 
\end{equation}

\noindent represent the color component for the $n$-th voxel as viewed by the $N_c$ camera rigs. For this approach, the \gls{klt} is applied over the $\vec{c}(n)$ vector signal as 

\begin{equation}\label{fun:color-comp}
    \vec{s}(n) = \matr{H}_c\vec{c}(n) , 
\end{equation}

\noindent
where $\vec{s}(n) = [S_{n}^{1} , S_{n}^{2} , \dotsm , S_{n}^{N_c}]^T$ is the vector with the transformed coefficients for the $n$-th voxel. $\matr{H}_c$ is the $N_{c}\times N_{c}$ KLT matrix of the color channel \textit{C}, i.e. an orthogonal matrix made of the eigenvectors of the covariance matrix $\matr{K}_{cc}$ whose entries are $K_{ij} = \{ E [ ( C_n^i - E [ C_n^i ] ) (  C_n^j - E [ C_n^j ] ) ]  \}$. Each of the $S_n^{k}$ is an attribute of the $n$-th transformed voxel and the set of $\{S_n^{k}\}$ is a \gls{pc} to be transformed and encoded. The set $\{S_{n}^{1}\}$ is often called the DC coefficient \gls{pc}. The others ($\{S_{n}^{k > 1}\}$) are referred as AC \gls{pc}s. The AC coefficients may contain negative values and, because of that, they are made positive by adding an offset.
Moreover, since we allocate distortion instead of rate, all KLT channels for each of YUV are properly scaled and subject to the same QP. Therefore, the impact of our solution on the overall bit-rate consists of the transmission of both the DC and the AC coefficients, in addition to the point cloud's main color. The coefficients of $\matr{K}_{cc}$ can be incorporated in the high-level syntax of \gls{gpcc}’s bitstream, as well as a single-bit flag signaling the encoding with the plenoptic enhancement. These coefficients can be conveyed in the bitstream using 32 bits floating-point numbers. Since the covariance matrix is symmetrical, $\frac{N_c(N_c+1)}{2}$ coefficients for each color channel have to be transmitted. According to our experiments, this side information corresponds to less than 0.07\% of the total bit-rate.

We propose to encode the $\{S_n^k\}$, which are still redundant signals, using state-of-the-art attribute coding schemes such as \gls{raht} or \gls{lod}, which are used in MPEG \gls{gpcc}. 
The operations and functionality of those \gls{pcc} methods (\gls{raht} and \gls{lod}) can be found elsewhere \cite{emerging_mpeg, gpcc_codec_description, lift_description}.
In summary, however, the main advantage of using these methods for encoding the coefficients over what is used in \cite{sandri_ppc} stems from the ability to provide an intra-frame prediction feature that explores attribute correlation among neighboring voxels, achieving further bit-rate savings.

\begin{figure}[htb]
\centering
\includegraphics[width=0.45\textwidth]{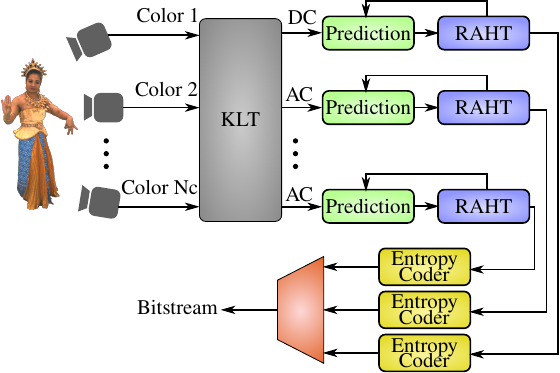}
\caption{Compression scheme for the intra-frame-predicted RAHT (proposed method 1).}
\label{fig:uraht-scheme} 
\end{figure}

\begin{figure}[htb]
\centering
\includegraphics[width=0.4\textwidth]{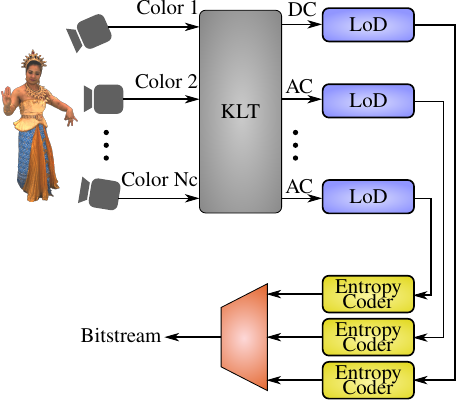}
\caption{Compression scheme using \gls{lod} (proposed method 2).}
\label{fig:lt-scheme} 
\end{figure}

In essence, we run $N_c$ attribute coders in parallel as depicted in Figs. \ref{fig:uraht-scheme} and \ref{fig:lt-scheme}.
Figure \ref{fig:uraht-scheme} illustrates the encoder based on \gls{raht}, wherein the $N_c$ \gls{klt} output signals are fed to the multiple encoders.   
Note that it is important, however, that we use intra-frame-prediction within \gls{raht} \cite{uraht_description, andre_icip} in order to further remove redundancy and to improve compression by exploiting the correlation between neighboring voxels. 
We have tested the use of intra-frame-prediction in both the DC and AC channels and the results  consistently pointed to a superior performance when using the prediction, which is described in \cite{uraht_description}.
A very similar approach based on LoD is illustrated in Fig. \ref{fig:lt-scheme} and is referred as Proposed Method 2.
In Figs. \ref{fig:uraht-scheme} and \ref{fig:lt-scheme}, we used the G-PCC entropy coder.

The intra-frame prediction for both of the proposed schemes allow the compression of the multiple attributes of \gls{ppc}s leveraging different characteristics. These are evaluated quantitatively -- both against each other, but also agains the state-of-the-art -- in the following section.


\section{Evaluation} \label{sec:results}

    
The experiments in this work were carried with version 10.0 of the \gls{gpcc} reference software (Test Model Categories 1 and 3 or TMC13) \cite{gpcc_codec_description}. Both our proposed coding schemes were compared to the works of Sandri et al. \cite{sandri_ppc} and Li et al. (MV-HEVC) \cite{lili}. We compared our solutions to \cite{lili}, which is a state-of-the-art but not compliant with V-PCC. Other methods that were compliant with either \gls{vpcc} or \gls{gpcc} did not prove competitive. Ours, however, is competitive and compliant with MPEG’s TMC13 (\gls{gpcc}).

We ran the experiments with the publicly available 8i Voxelized Surface Light Field Dataset (8iVSLF) \cite{jpeg_pleno_database}. These are the sequences being considered for the common test conditions (CTC) with respect to the MPEG efforts for plenoptic point cloud compression \cite{ctc_pleno_vpcc}, and were the basis for the benchmark experiments due to compliance with MPEG's standards. The creation of new datasets which explore additional characteristics of the PPCs for a more thorough evaluation is something that we are aware of, and it is regarded as future work. Each \gls{ppc} from this set has $N_c$ RGB color values that are associated with each different camera viewpoint, and the geometry information of the voxels is constrained within a cube of $4096 \times 4096 \times 4096$ voxels, whose resolution is often referred to as a \gls{pc} of ``depth-12". However, downsampled versions of the \gls{ppc}s from this dataset have also been generated which, among other advantages, provide reduced processing time and reduced memory consumption \cite{downsampled_camilo}. Therefore, tests for our proposed methods and the method in \cite{sandri_ppc} were also performed with ``depth-10" versions of those sequences. Details of the \gls{ppc}s are provided in Table \ref{tab:ppcs}.

\begin{table}[ht!]
\centering
\caption{Characteristics of the 8iVSLF dataset}
\label{tab:ppcs}
\begin{tabular}{cccc}
\hline
\multirow{2}{*}{\textbf{Sequence}} & \multicolumn{2}{c}{\textbf{Voxels}}   & \multirow{2}{*}{\textbf{$N_c$}} \\
                                   & \textbf{Depth 12} & \textbf{Depth 10} &                                 \\ \hline
\textit{Boxer}                     & 3493085           & 995099            & 13                              \\
\textit{Longdress}                 & 3096122           & 912518            & 12                              \\
\textit{Loot}                      & 3017285           & 869565            & 13                              \\
\textit{Redandblack}               & 2770567           & 839315            & 12                              \\
\textit{Soldier}                   & 4001754           & 1193515           & 13                              \\
\textit{Thaidancer}                & 3130215           & 283206            & 13                              \\ \hline
\end{tabular}
\end{table}

We calculate the distortion by considering the \gls{psnr} in between original and reconstructed Y channels for the $N_c$ cameras either concatenated as a single signal or taking the average across all cameras, in accordance to \cite{lili}. We use the former method except when comparing to MV-HEVC, in which case we use the latter. 


First, we compare our proposed methods 1 and 2. Every \gls{ppc} from the 8iVSLF dataset comes with a main RGB channel in addition to the $N_c$ colors from the different camera viewpoints, which is generated either by a weighted average of the $N_c$ colors or by taking the frontal view of the \gls{pc} \cite{8iVSLF}. Hence, we applied both coders over the non-transformed main color information in order to assess their performances in addition to our proposed compression schemes. In order to adequately compare the plenoptic results with the ones using the main color, both the DC and ACs coefficients in our proposed solution are scaled to fit an 8-bit representation.
Table \ref{tab:RAHT-LT-Main-Plenoptic} presents the comparison results for the six depth-10 8iVSLF sequences. \gls{bdpsnr} \cite{bd_metrics} results with the main color show an even comparison between coders. However, plenoptic results for the proposed method 1 present an average \gls{bdpsnr} value of $0.45$ dB over the method 2.

\begin{table}[h]
\centering
\caption{\gls{bdpsnr} (in dB) for the Y component of depth-10 \gls{pc}s, comparing Proposed Method 1 over 2.}
\label{tab:RAHT-LT-Main-Plenoptic}
\resizebox{\linewidth}{!}{%
\begin{tabular}{crcrrrr}
\hline
\textbf{PC}          & \textbf{Main} & \multicolumn{1}{r}{\textbf{Plenoptic}} &  & \textbf{DC} & \textbf{AC 1} & \textbf{AC2} \\ \cline{1-3} \cline{5-7} 
\textit{Boxer}       & $-0.02$                        & $0.27 $                                                 &  & $0.02$                       & $0.29 $                        & $0.28 $                       \\
\textit{Longdress}   & $0.10$                         & $0.72$                                                  &  & $0.26$                       & $0.26$                         & $0.33$                        \\
\textit{Loot}        & $-0.30$                        & $0.24 $                                                 &  & $0.02 $                      & $0.28 $                        & $0.32 $                       \\
\textit{Redandblack} & $0.07$                         & $0.29 $                                                 &  & $ 0.05$                      & $0.17 $                        & $0.21$                        \\
\textit{Soldier}     & $0.02$                         & $0.34 $                                                 &  & $-0.07 $                     & $0.17 $                        & $0.20$                        \\
\textit{Thaidancer}  & $0.04$                         & $0.81$                                                  &  & $0.37$                       & $0.21 $                        & $0.22$                        \\ \cline{1-3} \cline{5-7} 
\textit{Average}     & $-0.02$                        & $0.45$                                                  &  & $0.11$                       & $0.23$                         & $0.26 $
\\ \hline
\end{tabular}}
\end{table}




  In order to investigate where the gains come from, we broke down the results of the \gls{klt}-transformed color vectors by the \gls{rd} performance of the individual DC and AC coefficient \gls{pc}s. Some of the results are shown in Table \ref{tab:RAHT-LT-Main-Plenoptic}, in which the DC yields an average \gls{bdpsnr} gain of $0.11$ dB for the method 1 over the proposed method 2, and the first and second ACs yield higher gains of $0.23$ dB and $0.26$ dB, respectively. We believe that the prediction model in predictive RAHT performs better than \gls{lod}'s scheme for the transformed color information due to the selection of neighboring voxels. The \gls{lod} generation defines the order in which the colors are encoded. This order establishes which attribute values are available as references for prediction, which is based on the $k$-nearest-neighbors algorithm and uses point-to-point Euclidean distance thresholds. Therefore, the encoding order for this approach may not be optimal when using the transformed attributes. Hence, the lower correlation between neighbors for the AC-coefficient \gls{pc}s may produce larger residuals in method 2 in comparison to method 1. This may lead to worse coding performance over the scaled plenoptic colors. The usage of unscaled coefficients for both methods further accentuates these differences, as can be seen in Table \ref{tab:depth12-10}.

\begin{figure}[htb]
\centering
\includegraphics[width=0.45\textwidth]{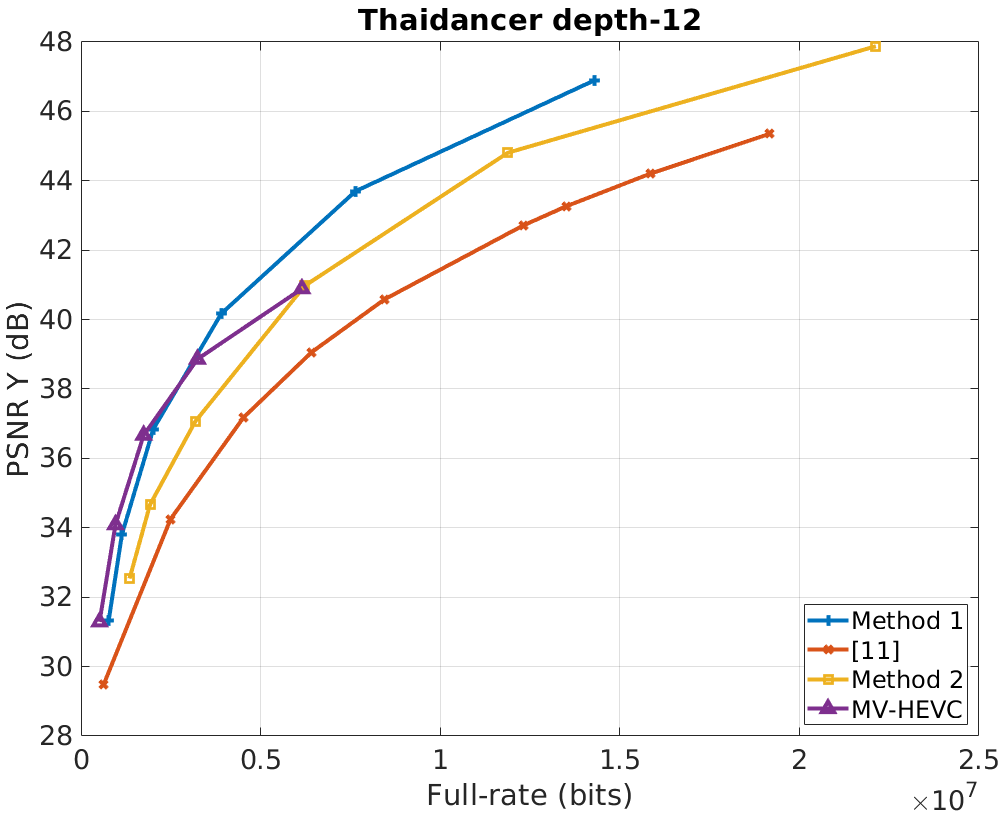}
\caption{Rate-distortion performance over the depth-12 \textit{Thaidancer} for the four compression schemes considered.}
\label{fig:depth12-thai} 
\end{figure}
\begin{figure}[thb]
\centering
\includegraphics[width=0.45\textwidth]{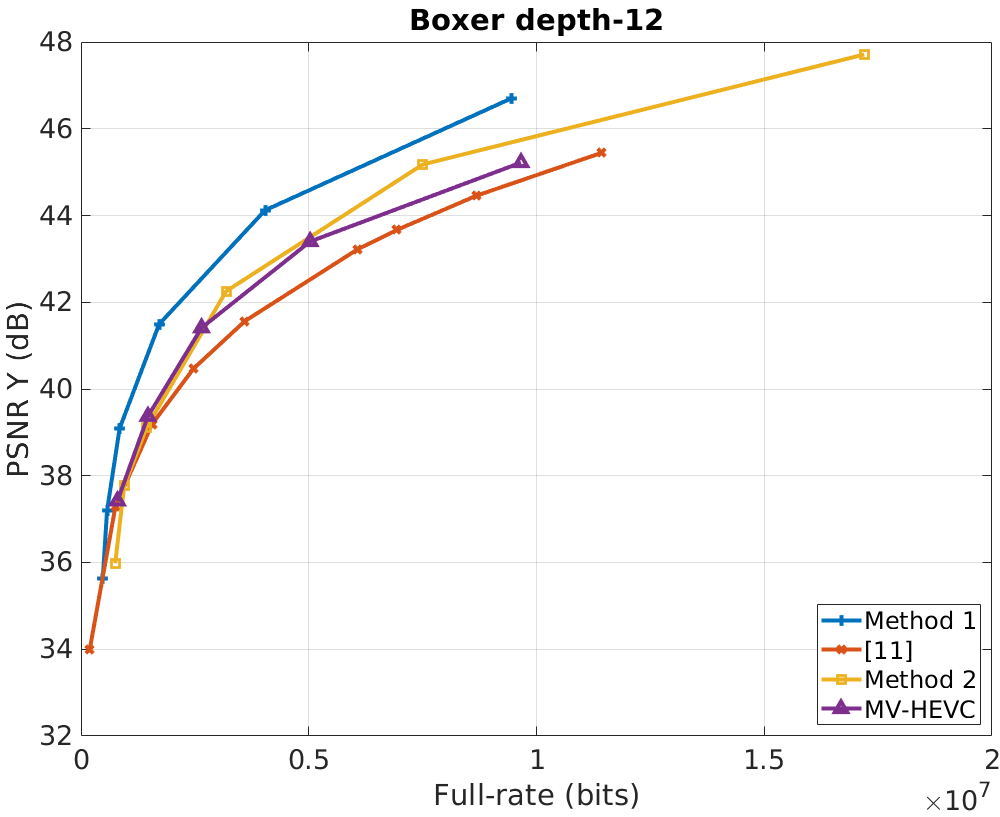}
\caption{Rate-distortion performance over the depth-12 \textit{Boxer} for the four compression schemes considered.}
\label{fig:depth12-boxer} 
\end{figure}

Table \ref{tab:depth12-10} presents the \gls{bdpsnr} \cite{bd_metrics} comparing our proposed methods 1 and 2 and MV-HEVC, using \cite{sandri_ppc} as an anchor. The BD metrics were computed using the points within the bit-rate range that comprises \gls{raht}'s quantization stepsizes from $12$ to $300$. The proposed method 1 significantly outperforms the others for most point clouds. Figs. \ref{fig:depth12-thai} and \ref{fig:depth12-boxer} present \gls{rd} curves comparing the four methods for the depth-12 \textit{Thaidancer} and \textit{Boxer} \gls{pc}s. Results show greater rate-distortion gains for method 1 in comparison to MV-HEVC for medium and high bit-rate cases.

\addtolength{\tabcolsep}{-2pt}    
\begin{table}[h]
\centering
\caption{\gls{bdpsnr}(in dB) comparisons of the Y component for different methods, using \cite{sandri_ppc} as reference.}
\label{tab:depth12-10}
\begin{adjustbox}{width=\linewidth,center}
\begin{tabular}{cccc|cc}
\hline
\multirow{2}{*}{\textbf{Sequence}}    & \multicolumn{3}{c}{\textbf{Depth 12}}                                                                           & \multicolumn{2}{c}{\textbf{Depth 10}}            \\ \cline{2-6}
                                      & \textbf{MV-HEVC} & \textbf{Method 1} & \textbf{Method 2} & \textbf{Method 1} & \textbf{Method 2} \\ \hline
\textit{Boxer}       & 0.55                              & \textbf{1.78}      & 0.69                                 & \textbf{1.83}                                 & 0.95                                 \\
\textit{Longdress}   & 3.40                              &\textbf{3.42}      & 2.17                                 & \textbf{2.69}                                 & 1.79                                 \\
\textit{Loot}        & 1.88                              & \textbf{2.17}       & 1.11                                 & \textbf{1.82}                                 & 0.99                                 \\
\textit{Redandblack} & 1.66                              & \textbf{2.73}       & 1.66                                 & \textbf{2.36}                                 & 1.58                                 \\
\textit{Soldier}     & 2.13                              & \textbf{2.53}       & 1.54                                 & \textbf{1.99}                                 & 1.14                                 \\
\textit{Thaidancer}  & \textbf{3.37}    & 3.33                                 & 1.86                                 & \textbf{2.46}                                 & 1.27                                 \\ \hline
\end{tabular}
\end{adjustbox}
\end{table}
\addtolength{\tabcolsep}{2pt}

\begin{figure}[thb]
\centering
\includegraphics[width=0.45\textwidth]{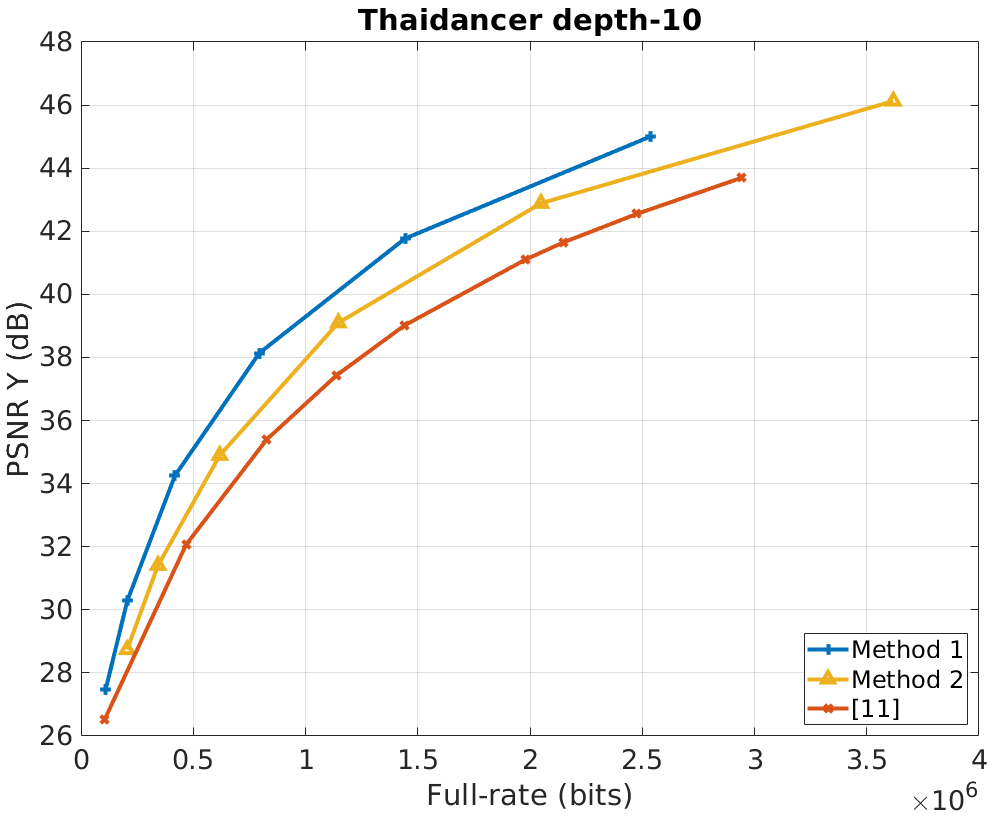}
\caption{Rate-distortion performance over the depth-10 \textit{Thaidancer} for the methods 1 and 2 compared to \cite{sandri_ppc}.}
\label{fig:depth10-redandblack} 
\end{figure}

Table \ref{tab:depth12-10} also presents the \gls{bdpsnr} results for both our proposed solutions compared to \cite{sandri_ppc} for the depth-10 \gls{ppc}s. As is the case with the depth-12 sequences, method 1 offers substantial coding gains in comparison to the method in \cite{sandri_ppc}. Fig. \ref{fig:depth10-redandblack} presents the \gls{rd} performance for the \textit{Thaidancer} sequence, which exhibits the same pattern found with the other \gls{ppc}s tested. Please note that MV-HEVC results are only available for depth-12 \gls{ppc}s.



\section{Conclusions} \label{sec:conclusions}
This work proposes the incorporation of plenoptic capability for MPEG's geometry-based encoder (\gls{gpcc}). This is achieved by compressing plenoptic point clouds using a combination of the \gls{klt} over the color vector of the different camera viewpoints followed by the usage of \gls{gpcc}'s attribute coders based on \gls{raht} and \gls{lod}.  
Results show that the \gls{raht}-based coding scheme (proposed method 1) provides substantial coding gains in comparison to competing approaches for both depth-12 and depth-10 \gls{ppc}s. It even surpasses a V-PCC-based approach, making this proposed solution the state of the art in the compression of \gls{ppc}s.

\FloatBarrier
\small
\bibliographystyle{./references/IEEEbib}
\bibliography{./references/references}

\end{document}

%% file: other/acronyms.tex
\newacronym{pc}{PC}{point cloud}
\newacronym{pcs}{PCs}{Point clouds}
\newacronym{ppc}{PPC}{plenoptic point cloud}
\newacronym{pcc}{PCC}{point cloud compression}
\newacronym{rd}{RD}{rate-distortion}

\newacronym{bdrate}{BD-Rate}{Bj{\o}ntegaard-delta rate}
\newacronym{bdpsnr}{BD-PSNR}{Bj{\o}ntegaard-delta PSNR}

\newacronym{klt}{KLT}{Karhunen-Lo\`{e}ve transform}
\newacronym{raht}{RAHT}{region-adaptive hierarchical transform}
\newacronym{uraht}{URAHT}{intra-frame-predicted RAHT}
\newacronym{lt}{LT}{lifting transform}

\newacronym{mpeg}{MPEG}{Moving Picture Expert Group}
\newacronym{gpcc}{G-PCC}{geometry-based point cloud compression}
\newacronym{vpcc}{V-PCC}{video-based point cloud compression}

\newacronym{8i}{8iVSLF}{8i Voxelized Surface Light Field Dataset}
\newacronym{psnr}{PSNR}{peak signal-to-noise ratio}
\newacronym{mse}{MSE}{Mean Square Error}
\newacronym{ctc}{CTC}{Common Test Conditions}
\newacronym{rgb}{RGB}{red-green-blue} 
\newacronym{lod}{LoD}{level of detail}

\newacronym{3d}{3D}{three-dimensional}
\newacronym{mv-hevc}{MV-HEVC}{Multiview High Efficiency Video
Coding}